\documentclass[runningheads,a4paper,amsmath,amssymb]{llncs}

\usepackage{amssymb}
\usepackage{graphicx}

\usepackage{url}
\usepackage{graphicx}%
\usepackage{dcolumn}%
\usepackage{bm}%
\usepackage{comment}
\usepackage{hyperref}%
\usepackage{amsmath}
\usepackage[usenames,dvipsnames]{color}
\usepackage{comment}
\usepackage[normalem]{ulem}
\usepackage{soul,xcolor}

\urldef{\mailsa}\path|aleeds@kth.se|
\newcommand{\keywords}[1]{\par\addvspace\baselineskip
\noindent\keywordname\enspace\ignorespaces#1}

\begin{document}
\setstcolor{red}
\mainmatter  %

\title{Flexoelectricity and Flexomagnetism }

\author{Alexander Edstr\"om\inst{1} \and
        Massimiliano Stengel\inst{2,3} }
\authorrunning{A. Edstr\"om and M. Stengel}
\institute{Department of Applied Physics, KTH Royal Institute of Technology,
           SE-11419 Stockholm, Sweden\\
           \email{aleeds@kth.se}
\and
           Institut de Ci\`encia de Materials de Barcelona (ICMAB-CSIC),
           Campus UAB, 08193 Bellaterra, Spain\\
           \email{mstengel@icmab.es}
\and
           ICREA -- Instituci\'o Catalana de Recerca i Estudis Avan\c{c}ats,
           08010 Barcelona, Spain}

\maketitle

\begin{abstract}
Strain gradients, unlike homogeneous strain, are odd under space inversion and unlock couplings that are strictly forbidden in the parent crystal. At the nanoscale, and in curved low-dimensional systems in particular, such gradients are readily attainable and often unavoidable, so that curvature acts as a symmetry-breaking field written into the lattice itself. This chapter reviews the resulting couplings to electric polarization and magnetism. After establishing the geometric relation between strain gradients and curvature, we review flexoelectricity as the better-established reference case: the bulk, surface and dynamic contributions that complicate its definition, its microscopic mechanisms, its manifestation in two-dimensional materials, and the first-principles methodology now available for its calculation. Using the magnetoelectric effect as a conceptual bridge, we then develop flexomagnetism in a broad sense, covering any coupling phenomena between magnetism and strain gradients, and specifically discuss the linear magnetization response to curvature and curvature-induced Dzyaloshinskii–Moriya interactions, beyond the established curvilinear treatment. Finally, we briefly describe some additional effects fitting under this broad umbrella of flexomagnetism, including nonlinear ones, inhomogeneous magnetoelectric and Rashba-type phenomena, and connections to altermagnetism.
\end{abstract}

\section{Introduction}\label{sec.intro}

Bend a crystal, and it becomes polar. %
A homogeneous strain, no matter how large, may stretch bonds and lift rotational
symmetries, but it cannot remove an inversion center: the strain
tensor is even under parity and piezoelectricity is forbidden in materials with inversion center (see Table~\ref{table.sym}). 
A strain gradient suffers no such restriction: being a parity-odd 
tensor,
it unlocks couplings that are strictly
forbidden in the parent crystal. 
At macroscopic dimensions this observation is of little
consequence, since attainable strain gradients scale inversely with the
sample size. 
At the nanoscale the situation is reversed: linear gradients of
$10^{6}~\mathrm{m}^{-1}$ have been engineered in epitaxial perovskite
nanostructure arrays~\cite{Tang2017}, and bent freestanding membranes
sustain upwards of $10^{7}~\mathrm{m}^{-1}$ uniformly across their thickness~\cite{Cai2022} — {more than enough to produce observable
effects~\cite{Flexoel_Zubkoetal,Cross2006}.}

In low-dimensional
systems, curvature is not merely attainable but often
unavoidable. The intrinsic ripples of free-standing
two-dimensional crystals~\cite{Fasolino2007} are a prime example, while curved and
three-dimensional architectures have become a deliberate design element of
modern nanomagnetism~\cite{Streubel_2016,Pacheco2017,Sheka2021,curvilinmicromag2022}.
Gradients of strain, in short, have been promoted from a parasitic
by-product of imperfect samples to a symmetry-breaking field in their own right. 
Unlike  electric or magnetic fields, which are typically sourced from outside, strain is often written into the lattice itself by geometry, growth, or patterning, and confined with a locality that stray fields cannot match.

The best known physical manifestation of spatially inhomogeneous deformations 
is flexoelectricity: the electric polarization induced linearly by a strain
gradient~\cite{Kogan1964,PhysRevB.34.5883,Tagantsev2015}. 
Its defining
virtue is universality: the corresponding fourth-rank tensor is allowed in
every crystal class, so that \emph{any} insulator polarizes when bent.
The pioneering experimental demonstration of a large response in
high-permittivity dielectrics~\cite{Ma_Cross_2001}
was soon followed by the exploration of ever more ambitious functionalities,
ranging from polarization rotation~\cite{Catalan2011} and writing of ferroelectric domains~\cite{Lu59} to silicon-compatible actuation~\cite{Bhaskar2016} and photoflexoelectricity~\cite{Shu2020}.
From the theory point of view, the definition of flexoelectricity is riddled with subtleties (regarding, e.g., the respective roles of bulk versus surface
contributions, static versus dynamic gradients, etc.~\cite{PhysRevB.34.5883}) to the point that the very bulk nature of the effect has long been debated~\cite{PhysRevLett.105.127601}.
In recent years, many of these formal puzzles have been solved, and 
a comprehensive first-principles methodology to calculate
flexoelectric coefficients now exists as part of
the publicly available software package ABINIT~\cite{abinit2020,abinit2025}. 
This achievement was a long time coming. After Tagantsev's seminal work ~\cite{PhysRevB.34.5883}, many years passed before the first ab initio attempts~\cite{PhysRevLett.105.127601,Hong_2010}.
Methodological progress then accelerated rapidly~\cite{PhysRevB.90.201112}, although several technical and conceptual gaps~\cite{PhysRevB.98.125133,PhysRevB.98.075153,PhysRevB.99.085107} had to be addressed before arriving at the present formulation~\cite{PhysRevX.9.021050,PhysRevB.105.064101}.
We close this brief overview by emphasizing the importance of atomically thin 2D materials, both in the context of the theory and prospective applications. Indeed, their flexural phonon modes provide a natural platform to study curvature effects that are free from the aforementioned ambiguities, and can be easily accessed experimentally thanks to their low bending modulus. A theory and computational methodology that deals with the specifics of 2D systems now also exists~\cite{PhysRevLett.127.216801,PhysRevLett.131.236203}, nicely complementing the earlier work on 3D crystals.

The magnetic case is less explored, and here time-reversal symmetry changes the rules of the game (Table~\ref{table.sym}). 
Since magnetization is a time-odd axial vector while a strain gradient is a time-even polar tensor, the direct analogue of flexoelectricity is far from universal: it is allowed only in certain magnetic point groups.
Like piezomagnetism, it requires magnetic order, but transforms differently under space inversion;
the relevant crystal classes are closely related to those hosting the linear magnetoelectric (ME) effect~\cite{PhysRevB.84.174112,ME_Dzyaloshinsky,ME_Astrov1960,ME_Astrov,Fiebig_2005}.
In this sense, flexomagnetism is a sensitive probe of magnetic symmetry, and has only recently been detected experimentally~\cite{Makushko2022}.
Universality is, however, recovered in a different guise: because a strain
gradient breaks inversion symmetry in \emph{every} magnet, it always
induces a Dzyaloshinskii--Moriya
interaction~\cite{DZYALOSHINSKY1958241,PhysRev.120.91,PhysRevB.98.214414},
with curvature-driven chiral and topological textures as its most striking
consequence~\cite{PhysRevLett.112.257203,Sheka_2015,PhysRevLett.114.197204,PhysRevLett.123.077201}.
Together with second-order effects on the magnetic
anisotropy~\cite{PhysRevLett.128.177202},
curvature-induced Rashba fields~\cite{gentile1,gentile2,Manchon2015},
and emerging connections to altermagnetism and $p$-wave
magnetism~\cite{PhysRevX.12.040501,PhysRevLett.134.116701,hellenes2024pwavemagnets,jin2026rolling2d_pwave},
these findings have motivated a broader reading of ``flexomagnetism'' as the family of all couplings involving magnetism and strain gradients.

\begin{table}[h]
\centering
\caption{Symmetry properties of piezo-, flexo-, and magnetoelectric tensors.}
\label{table.sym}
\begin{tabular}{@{}lccccc@{}}
\hline
\textbf{Tensor} & \textbf{Rank} & \textbf{Type} & \textbf{$I$} & \textbf{$T$} & \textbf{$IT$ ($I^\prime$)} \\ %
Piezoelectric   & 3rd & Polar & odd  & even & odd   \\ 
Flexoelectric   & 4th & Polar & even & even & even   \\ %
Magnetoelectric & 2nd & Axial & odd  & odd  & even   \\ %
Piezomagnetic   & 3rd & Axial & even & odd  & odd    \\
Flexomagnetic   & 4th & Axial & odd  & odd  & even    \\
\end{tabular}
\end{table}

This chapter is organized accordingly. Section~\ref{sec.curv} establishes
the geometric relation between strain gradients and curvature, fixing the
notation used throughout. Section~\ref{sec.flexoel} reviews
flexoelectricity as the better-established reference case: the bulk effect and the surface and dynamic subtleties that complicate its definition
(Secs.~\ref{sec.surf_FxE} and \ref{sec.dyn_FxE}); its microscopic
mechanisms (Sec.~\ref{sec.flexoel_mech}) and their
manifestation in two-dimensional materials (Sec.~\ref{sec.flexoel_2d}); 
and the underlying first-principles theory
 (Sec.~\ref{sec.flexoel_theory}).
In Section~\ref{sec.flexomag} we discuss flexomagnetism in the broader sense. 
First, we bridge the two halves of the chapter by reviewing models of magnetism and the basics of the magnetoelectric effect (Sec.~\ref{sec.mag_bg}).
Next, we discuss the linear magnetization response to
curvature (Sec.~\ref{sec.lin_flexomag}) and curvature-induced
Dzyaloshinskii--Moriya interaction (DMI) beyond the standard curvilinear
treatment (Sec.~\ref{sec.intrinsicDMI}). 
Finally, we briefly discuss effects that are nonlinear in curvature, and other related couplings between spins and gradients in Sec.~\ref{sec.otherFxM}, before presenting outlook and conclusions in Sec.~\ref{sec.concl}.

\section{Preliminaries: Strain Gradients and Curvature}\label{sec.curv}

In three-dimensional (3D) crystals, a general mechanical deformation is described by a $\mathbb{R}^3 \rightarrow \mathbb{R}^3$ mapping $\mathbf{r}^\prime = \mathbf{r} + \mathbf{u}$, where $\mathbf{u}$ is the deformation field with components $u_\alpha$. The symmetric strain tensor %
\begin{equation}\label{eq.strain}
    \varepsilon_{\alpha\beta} = \frac{1}{2}\left( u_{\alpha,\beta} + u_{\beta,\alpha} \right)    
\end{equation}
is the symmetrized form of the strain $u_{\alpha,\beta} = \partial_\beta u_\alpha$. 
Flexoelectricity and flexomagnetism are the couplings of electric polarization, and magnetism, respectively, to derivatives of strain, i.e. to strain gradients 
\begin{equation}\label{eq.straingradI}
    u_{\alpha,\beta\gamma} = \frac{\partial^2 u_{\alpha}}{ \partial r_\beta \partial r_\gamma}. 
\end{equation}
The form in Eq.~\ref{eq.straingradI} is usually referred to as the type I representation of the strain gradient~\cite{Stengel_Vanderbilt}. The type II strain gradient tensor instead is the gradient of the symmetrized strain %
\begin{equation}\label{eq.straingradII}
\varepsilon_{\alpha\beta,\gamma} = \frac{\partial \varepsilon_{\alpha\beta}}{ \partial r_\gamma} = \varepsilon_{\beta\alpha,\gamma}.
\end{equation}
In both representations, the strain gradient is a rank three polar tensor, symmetric in two of the indices; they can be related by $\varepsilon_{\alpha\beta,\gamma} = \tfrac{1}{2}(u_{\alpha,\beta\gamma}+u_{\beta,\alpha\gamma})$ and $u_{\alpha,\beta\gamma} = \varepsilon_{\alpha\beta,\gamma}+\varepsilon_{\alpha\gamma,\beta}-\varepsilon_{\beta\gamma,\alpha}$. If nothing else is specified we will use type II strain gradients (Eq.~\ref{eq.straingradII}) in the following. 

Alternatively, we can use the deformation gradient $F_{\alpha\beta} = \partial_\beta r_\alpha^\prime =\delta_{\alpha\beta} + u_{\alpha,\beta}$, to write the metric tensor $g_{\alpha\beta} = F_{\gamma\alpha}F_{\gamma\beta}$ and the strain tensor as the linearized Lagrange strain tensor $\varepsilon_{\alpha\beta} \approx \frac{1}{2} \left( g_{\alpha\beta} - \delta_{\alpha\beta} \right)$. Then the type II strain gradient can be computed as half the gradient of the metric, while the type I strain gradient is the Christoffel symbol 
\begin{equation}
    \Gamma_{\alpha\beta\gamma} = \frac{\partial^2 r^\prime_\delta}{\partial r_\beta \partial r_\gamma} \frac{\partial r^\prime_\delta}{\partial r_\alpha } 
\end{equation}
in the linear regime.

{The deformation of a two-dimensional object can also be framed in the 
language of differential geometry, with the main difference that the mapping
is now of the type $\mathbb{R}^2 \rightarrow \mathbb{R}^3$.
Thus, the deformation gradient now becomes a $3\times 2$ matrix, which we can conveniently express in terms of two independent tangent vectors $\mathbf{r}_\alpha = \partial \mathbf{r} / \partial \xi_\alpha$. ($\xi^\alpha = u,v$ are the curvilinear coordinates that parametrize the surface.) The metric, known
as \emph{first fundamental form} in the context of surfaces,
is a $2 \times 2$ tensor, $g_{\alpha \beta} = \mathbf{r}_\alpha \cdot \mathbf{r}_\beta$: it provides a measure of the local in-plane strain.
The metric and its gradients, however, do not provide information about the curvature of the embedded surface: to that end, it is useful to introduce 
the second fundamental form~\cite{doCarmo2016},}
\begin{equation}\label{eq.curvtensor}
S_{\alpha\beta} = S_{\beta\alpha} = \frac{\partial^2 r^\gamma}{\partial \xi^\alpha \partial \xi^\beta } n_\gamma \approx \Gamma_{z \alpha\beta}, %
\end{equation}
where $\mathbf{n} = \mathbf{r}_u \times \mathbf{r}_v / |\mathbf{r}_u \times \mathbf{r}_v| $ is the normal to the oriented surface at point ${\bf r}$.
The symmetric $2\times 2$ matrix $S_{\alpha\beta} \approx -\varepsilon_{\alpha\beta,z}$ transforms like a rank 2 tensor under in-plane rotations, but like the components $\varepsilon_{\alpha\beta,z}$ of the 3D strain gradient tensor under general 3D operations.  
The eigenvalues of ${\bf S}$ are known as principal curvatures ($K_{1,2}$), and together with the corresponding eigenvectors ($\bm{\lambda}_{1,2}$) they provide a complete description of the local flexural deformation. The mean and Gaussian curvatures are given by $(K_1+K_2)/2$ and by $G=K_1 K_2$ (the average and product of the principal curvatures), respectively, corresponding to half the trace and the determinant of the curvature tensor. The case where one of the two eigenvalues vanishes describes cylindrical bending,  
while a diagonal tensor with $K_1 = K_2$ locally describes a spherical curvature.

\section{Flexoelectricity}\label{sec.flexoel}

The flexoelectric effect can be described as the first derivative of the
polarization ${\bf P}$ with respect to the strain gradient tensor components $\varepsilon_{\beta\gamma,\lambda}$ via the fourth-rank flexoelectric tensor,
\begin{equation}\label{eq.flexoel_tensor}
    \mu_{\alpha\lambda,\beta\gamma}^{E} = \frac{\partial P_\alpha}{\partial \varepsilon_{\beta\gamma,\lambda}}.
\end{equation}
After its prediction by Kogan~\cite{Kogan1964} in the sixties, its study remained largely  
dormant for several decades, as it was believed
to be too small to be of practical interest. %
An important turning point came with the work of Cross~\cite{Cross2006}, experimentally measuring an unexpectedly large flexoelectric coefficient at room temperature. The reported values were as large as $\mu_{1122} = 4~\mu\mathrm{C/m}$, in the ceramic relaxor ferroelectric Pb(Mg$_{1/3}$Nb$_{2/3}$)O$_3$~\cite{Ma_Cross_2001}, and $\mu_{1122} = 100~\mu\mathrm{C/m}$ in paraelectric Ba$_{0.67}$Sr$_{0.33}$TiO$_3$~\cite{10.1063/1.1518559}. 
This finding was explained by observing that in both materials the dielectric constant is remarkably large~\cite{Ma_Cross_2001,10.1063/1.1518559}, and the flexoelectric effect is expected to scale accordingly~\cite{Tagantsev2015}. Cross and coworkers also measured the converse effect in Ba$_{0.67}$Sr$_{0.33}$TiO$_3$~\cite{Fu2006_flexoel} and extracted a similarly large value of $\mu_{1111} = 121~\mu\mathrm{C/m}$. 
The next important experimental development was the single crystal measurements on SrTiO$_3$, allowing all three flexoelectric coefficients to be determined to  $\mu_{1111} = 0.2~\mathrm{nC/m}$, $\mu_{1122} = 7~\mathrm{nC/m}$ and $\mu_{1212} = 5.8~\mathrm{nC/m}$~\cite{PhysRevLett.99.167601,PhysRevLett.100.199906}. %
More recent measurements on a number of single crystal oxides (${\mathrm{SrTiO}}_{3}, {\mathrm{KTaO}}_{3}, {\mathrm{TiO}}_{2}$, ${\mathrm{YAlO}}_{3}$) also obtained values of the order of few nC/m~\cite{PhysRevMaterials.6.055005}. 

In spite of the intense experimental activity, the unambiguous interpretation of these results in terms of well-defined microscopic mechanisms remains problematic. 
Indeed, the measured polarization response upon bending is known to contain a contribution from surface piezoelectricity that cannot be distinguished from
bulk flexoelectricity~\cite{PhysRevB.90.201112,TagantsevYurkov2012}: both effects display the same scaling with thickness. 
This theoretical prediction is supported by an experimental study of the flexoelectric anisotropy in BaTiO$_3$~\cite{PhysRevLett.115.037601}: the measured values are incompatible with their interpretation as a genuine bulk property.
This important aspect of the flexoelectric effect should be kept in mind when 
developing theoretical models, as we shall see in the following sections.

\subsection{Bulk Flexoelectricity}\label{sec.flexoel_bulk}

From the point of view of macroscopic theory, the electromechanical free energy density can be written as %
\begin{align}\label{eq.flexoPhi}
   \Phi = & \frac{\left({\chi}^{E}\right)^{-1}_{\alpha\beta} }{2} P_\alpha P_\beta + \frac{1}{2}c_{\alpha\beta\gamma\delta} \varepsilon_{\alpha\beta} \varepsilon_{\gamma\delta} - \vartheta_{\alpha\beta\gamma}P_\alpha \varepsilon_{\beta\gamma} + \frac{g_{\alpha\beta\gamma\delta}}{2} \frac{\partial P_\alpha}{\partial r_\beta} \frac{\partial P_\gamma}{\partial r_\delta} \nonumber \\
   & - \frac{f_{\alpha\beta\gamma\delta}}{2} \left( P_\alpha \frac{\partial \varepsilon_{\gamma\delta}}{\partial r_\beta} - \varepsilon_{\gamma\delta} \frac{\partial P_\alpha}{\partial r_\beta} \right) - P_\alpha E_\alpha - \varepsilon_{\alpha\beta} \sigma_{\alpha\beta},    
\end{align}
with electric susceptibility $\chi^{E}_{\alpha\beta}$, elastic tensor $c_{\alpha\beta\gamma\delta}$, piezoelectric tensor $\vartheta_{\alpha\beta\gamma}$ and flexocoupling tensor $f_{\alpha\beta\gamma\delta}$. 
Minimizing the free energy $\int \Phi \mathrm{d}V$ w.r.t. $P_\alpha$ leads to the constitutive equation for the electric polarization
\begin{equation}\label{eq.flexoel}
   P_\alpha = \chi^{E}_{\alpha\beta}E_\beta + e_{\alpha\beta\gamma} \varepsilon_{\beta\gamma} + \mu_{\alpha\beta\gamma\delta}^E \frac{\partial \varepsilon_{\gamma\delta}}{ \partial r_\beta}, 
\end{equation}
where the second and third terms represent piezoelectricity and flexoelectricity, that is the electric polarization induced linearly with strain and strain gradients, respectively. The piezoelectric and flexoelectric tensors are $e_{\alpha\beta\gamma} = \chi^{E}_{\alpha\eta} \vartheta_{\eta\beta\gamma}$ and $\mu_{\alpha\beta\gamma\delta}^E = \chi^{E}_{\alpha\eta} f_{\eta\beta\gamma\delta}$. 
Minimization of the free energy with respect to strain instead leads to the converse effect 
\begin{equation}
\sigma_{\alpha\beta} = c_{\alpha\beta\gamma\delta}\varepsilon_{\gamma\delta} 
- \vartheta_{\gamma\alpha\beta}P_\gamma
+ f_{\gamma\delta\alpha\beta} \frac{\partial P_\gamma}{\partial r_\delta} , 
\end{equation}
whereby a spatially varying electric polarization leads to a mechanical stress in the crystal. 

While the piezoelectric tensor $e_{\alpha\beta\gamma}$ is a polar rank 3 tensor, odd under space inversion symmetry, flexoelectricity is described by the rank 4 polar tensor $\mu_{\alpha\beta\gamma\delta}^E$, 
which is even under space inversion symmetry. Therefore, while crystal structures with inversion center do not permit the piezoelectric effect, the flexoelectric tensor has non-zero elements in all crystallographic point groups. 
\begin{figure}[h!]
	\centering
	\includegraphics[width=0.6\textwidth]{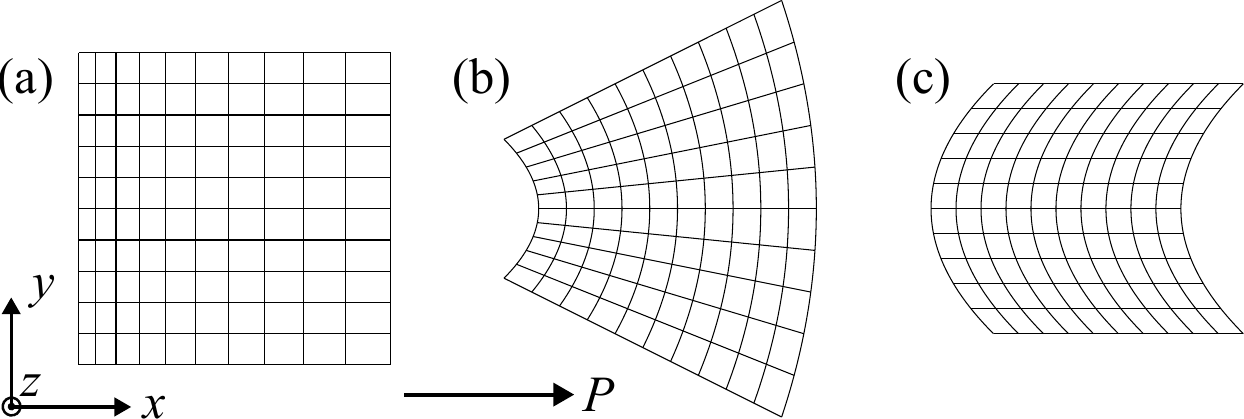}
	\caption{Schematic illustrations of the three strain gradients corresponding to the flexoelectric coefficients (a) $\mu_\mathrm{L} = \mu_{1111}^{E}$ (b) $\mu_\mathrm{T} = \mu_{1122}^{E}$ (c) $\mu_\mathrm{S} = \mu_{1212}^{E}$. With the coordinate system as in the figure, the electric polarization is along $\hat{x}$. 
    }
	\label{fig.struct}
\end{figure}
Being a fourth rank tensor, symmetric in two indices, the flexoelectric tensor can have at most 54 elements. In a high symmetry cubic crystal these are reduced to three independent, non-zero elements, $\mu_\mathrm{L} = \mu_{1111}^{E}$, $\mu_\mathrm{T} = \mu_{1122}^{E}$, $\mu_\mathrm{S} = \mu_{1212}^{E}$, for the ``longitudinal", ``transverse" and ``shear" components. These three components are non-zero in all materials.

\subsection{Surface Piezoelectricity}\label{sec.surf_FxE}

The description embodied in Eqs.~(\ref{eq.flexoPhi}) and (\ref{eq.flexoel})
tacitly assumes that $\mu^{E}_{\alpha\beta\gamma\delta}$ is a bulk materials constant on the
same footing as the piezoelectric tensor $e_{\alpha\beta\gamma}$. This assumption,
however, is considerably more delicate than it appears. A strain gradient,
unlike a homogeneous strain, cannot be sustained in a bounded sample
without inhomogeneously deforming its boundaries; and since any surface
locally breaks inversion symmetry, the surface regions of an otherwise
centrosymmetric crystal respond \emph{piezoelectrically} to this
deformation. The surface-induced dipole densities at opposite faces of a
bent plate carry opposite signs but unequal magnitudes, and their imbalance
is itself proportional to the applied strain gradient times thickness. As a result, the
surface contribution to the flexoelectric response of a finite sample does
\emph{not} vanish in the thermodynamic limit: it remains of the same order
as the static bulk effect, and even shares its scaling with the dielectric
permittivity~\cite{PhysRevB.34.5883,TagantsevYurkov2012,YurkovTagantsev2016}. %
This observation lies at the heart of the long-standing debate on whether
flexoelectricity can be defined as a bulk property at
all~\cite{PhysRevLett.105.127601,Flexoel_Zubkoetal,Yudin_2013}.
First-principles theory sharpens the statement: the open-circuit
flexovoltage of a bent film separates into a bulk and a surface term, each
individually dependent on the choice of reference (e.g., the treatment of
the surface truncation or of the microscopic current density), while only
their \emph{sum} constitutes a measurable
quantity~\cite{Stengel2013NatCommun,PhysRevB.90.201112,Stengel_Vanderbilt}. %
In SrTiO$_3$, remarkably, the surface term can be large enough to reverse
the sign of the total response~\cite{PhysRevB.90.201112}. The same
ambiguity is visible already at the level of Eq.~(\ref{eq.flexoPhi}):
augmenting $\Phi$ by a total derivative such as
$\partial_\delta (P_\gamma \varepsilon_{\alpha\beta})$ leaves the bulk Euler--Lagrange equations (and hence $\mu^{E}_{\alpha\beta\gamma\delta}$)
untouched but, upon integration by parts, transfers weight to the boundary.
The partition between the ``direct'' and ``converse'' forms of the flexocoupling 
is therefore a matter of convention
that acquires physical meaning only once the surfaces are
specified~\cite{Yudin_2013,Tagantsev2015}.

This has direct implications for the interpretation of the results
of a first-principles calculation of the flexoelectric tensor. 
Since the electrical boundary conditions are ill-defined at the bulk level, 
an arbitrary choice had to be made in the code (ABINIT) to fix the local electrostatic reference potential in an inhomogeneously strained crystal.
Out of computational convenience, such reference is set to the average electrostatic potential in the current ABINIT implementation.
Such a choice, however, does not correspond to realistic situations, where the boundary conditions are set by the sample boundary. This means that the calculated bulk flexoelectric coefficients have no physical significance of their own, and should be regarded as intermediate ingredients in the theoretical model.
As noted above, the reference ambiguity~\cite{PhysRevB.93.245107} only cancels out when the relevant surface terms are consistently incorporated in the model.
Two-dimensional suspended layers are an important case where the above complications disappear: these material systems are treated as a whole unit without any distinction between bulk and surface.
As such, the flexoelectric coefficients can be determined as well-defined 
properties of the crystal, unlike in the 3D case.

\subsection{Dynamic Flexoelectricity}\label{sec.dyn_FxE}
 
A second, independent obstacle to interpreting $\mu^{E}_{\alpha\beta\gamma\delta}$ as a unique
materials constant arises when one asks \emph{how} the strain gradient is
produced. A gradient may be imposed statically, e.g.\ by bending a beam, or
carried dynamically by a long-wavelength acoustic phonon. 
{In the former case, which is how the vast majority of 
the reported measurements have been performed, not all of the
components of the flexoelectric tensor can be accessed.
Indeed, a static inhomogeneous deformation is bound to satisfy
the mechanical equilibrium condition, stating that the force -- and hence the mechanical acceleration via Newton's law -- experienced by any material point
in the sample must vanish.
This equilibrium condition is simply formulated by requiring that the 
divergence of the local stress vanish,
\begin{equation}
    f({\bf r}) = \nabla \cdot \bm{\sigma}({\bf r}) = 0.
\end{equation}
Within the linear regime, the stress is related to the strain via 
the elastic tensor, which immediately leads to the following 
relation
\begin{equation}\label{eq.elast_equi}
 \sum_{\lambda\beta\gamma}   c_{\alpha\lambda,\beta\gamma} \varepsilon_{\beta\gamma,\lambda} = 0.
\end{equation}
This condition significantly reduces the linear combinations 
of strain gradient components that can be stabilized at any given point.
A simple example to appreciate this point is the transverse
strain-gradient deformation of Fig.~\ref{fig.struct}(b): since the cells at
the outer surface experience a (transverse) tensile strain, and those at the inner surface experience a (transverse) compressive strain, a \emph{longitudinal} stress gradient 
that is proportional to $\varepsilon_{22,1} c_{12}$ is present in the structure.
To compensate for this stress, a longitudinal strain gradient of the type in Fig.~\ref{fig.struct}(a) sets in, whose amplitude is proportional to $-\nu \varepsilon_{22,1}$ and $\nu=c_{12}/c_{11}$ is the two-dimensional counterpart of the Poisson ratio.

{So how did the experiment of Ref.~\cite{PhysRevLett.99.167601} achieve
the simultaneous measurement of the three independent flexoelectric 
coefficients in SrTiO$_3$? The answer is that two out of three values were obtained
via static measurements, and the system of linear equations was subsequently complemented with the results of a \emph{dynamic} measurement, thus making it possible to determine the three independent values. 
This procedure, while in principle justified, raises some puzzling questions about the interpretation of the results. 
For example, since we have mixed static and dynamic measurements, should we regard the coefficients that emerged from the linear system of equations as static or dynamic properties of the sample? 
This question is especially pressing in light of Tagantsev's distinction~\cite{PhysRevB.34.5883,Tagantsev2015} between static and dynamic contributions to the flexoelectric effect, suggesting the possibility that they could be probed separately in an experiment.}

{The first-principles theory of flexoelectricity that emerged in recent years managed to provide a conclusive answer.
Flexoelectricity is now understood as an inherently dynamical effect,
where all the individual coefficients of the tensor $\bm{\mu}^{E}$ contain an
explicit (and significant) dependence on the ionic masses~\cite{PhysRevB.88.174106}.
However, such a mass dependence cancels out in all linear combinations
of the coefficients that satisfy Eq.~\ref{eq.elast_equi}, consistent with the static nature 
of the effect under mechanical equilibrium conditions~\cite{PhysRevB.105.064101}.
Remarkably, this ambiguity disappears in all components of the flexoelectric tensor that correspond to a vanishing elastic restoring force: a notable example is the flexural deformation of a two-dimensional crystal or membrane under vanishing stress, where the flexoelectric effect is a genuine static property of the system~\cite{PhysRevLett.127.216801}.}

\begin{figure}[b!]
	\centering
	\includegraphics[width=0.6\textwidth]{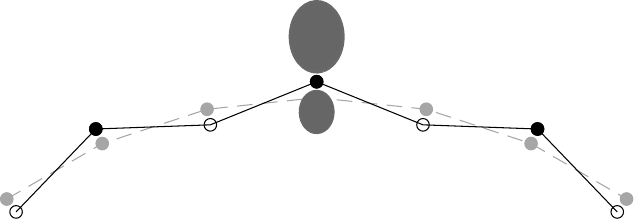}
	\caption{Schematic illustration of the flexoelectric contributions for a diatomic system. The lattice contribution occurs due to internal relaxation of the atoms upon applying strain gradient, as shown by the displacement of inequivalent black and white atoms being shifted from the positions of the gray atoms, representing positions after applying a strain gradient but no internal relaxation of the positions. The electronic contribution is caused by asymmetric distortion of the electronic charge density, as illustrated by the gray cloud around the atom in the center. }
        \label{fig.flexoel_chain}
\end{figure}

\subsection{Microscopic Mechanisms}\label{sec.flexoel_mech}

The microscopic mechanism of flexoelectricity can be divided into a
lattice-mediated and an electronic contribution, schematically illustrated
in Fig.~\ref{fig.flexoel_chain}, in close analogy with the familiar decomposition of the
piezoelectric response into clamped-ion and internal-strain
parts~\cite{Stengel_Vanderbilt,Flexoel_Zubkoetal}. The electronic
(clamped-ion, or ``frozen-ion'') part is defined by mapping the atomic
positions homogeneously onto the deformed configuration, without allowing
any additional relaxation: the induced polarization then stems entirely
from the quantum-mechanical redistribution of the electronic charge density
in response to the inhomogeneous
deformation~\cite{PhysRevLett.105.127601,PhysRevB.84.180101}. That this
contribution is irreducibly quantum mechanical is best appreciated by
noting that it survives in cases where any rigid-ion picture would predict
a vanishing response: elemental crystals such as silicon or diamond, whose
Born effective charges are identically zero, nonetheless possess a sizable
flexoelectric tensor, as do carbon nanostructures where curvature-induced
charge redistribution is the sole operative
mechanism~\cite{PhysRevB.77.033403,PhysRevB.84.180101}.

The lattice-mediated part results from the relaxation of the internal
degrees of freedom in the atomic positions once the constraint of a
homogeneous mapping is released. A strain gradient, unlike a homogeneous
strain, exerts a net force on each sublattice even in a centrosymmetric
crystal; the resulting internal displacements are polar, and the
polarization they carry is obtained by weighting them with the Born
effective charges. This is the mechanism identified in the early
lattice-dynamical treatments of
flexoelectricity~\cite{PhysRevB.34.5883,Tagantsev2015}, where it was
described via the first moments of the interatomic force constants within a
rigid-ion (point-charge) model. The modern long-wave formulation of
density-functional perturbation theory recovers both pieces on the same
footing, as successive orders of the charge-density and force response to a
long-wavelength acoustic phonon~\cite{PhysRevB.88.174106,PhysRevB.88.174107,PhysRevB.93.245107,PhysRevB.99.085107}. %
{The intimate relation between flexoelectricity and continuum elasticity that we hinted at earlier manifests itself in the form of exact sum rules on the flexoelectric force-response tensors~\cite{PhysRevB.105.064101}, which underpin the mechanical equilibrium condition of Eq.~\ref{eq.elast_equi}.}

{In 3D crystals it is hard to compare the relative weight of 
electronic and lattice-mediated contributions to the flexoelectric 
coefficients because of the reference potential ambiguity that we commented on 
earlier.
We can, however, attempt a meaningful assessment in 2D, where the polar response 
to a flexural deformation is a well-defined static property of the system.
We shall come back to this point in the following section.}

\subsection{Flexoelectricity in 2D Materials }\label{sec.flexoel_2d}

{Recall that the flexural deformation of a 2D crystal or membrane can be conveniently described by strain gradient components of the type $\varepsilon_{\beta\gamma,z}$, where $\beta\gamma$ are in plane, and $z$ is the out-of-plane direction; the three independent entries correspond to the elements of the second fundamental form and we can write the flexoelectric effect for a 2D crystal as 
\begin{equation}\label{eq.flexoel_2D}
P_\alpha = \mu_{\alpha, n}^{E} s_n
\end{equation}
where $s_n$ is Voigt notation ($s_1 = S_{xx}$, $s_2 = S_{yy}$, $s_3 = S_{xy}$) for the second fundamental form (Eq.~\ref{eq.curvtensor}). Here $P_\alpha$ denotes the dipole moment per unit area, so that $\mu_{\alpha, n}^{E}$ has units of charge.
%
With three independent polarization directions and three curvature components, 
up to nine coefficients can be active in the least symmetric system. 
Many of the most popular 2D systems have either a $D_{3d}$ (monolayer SnS$_2$, bilayer BN, etc.) or $D_{3h}$ (monolayer MoS$_2$, monolayer BN, etc.) point group; in such cases the number of independent components reduces to two and one, respectively.}

A flexoelectric response along the out-of-plane direction (indicated by $z$ hereafter) is allowed by symmetry in both crystal classes, and manifests itself as an open-circuit voltage $\Delta V = K \mu_{z}^E / \epsilon_0 $ that is proportional to twice the mean curvature, $K=K_1+K_2$~\cite{PhysRevLett.127.216801}. ($\mu_{z}^E = \mu_{z,1}^E = \mu_{z,2}^E$; $\mu_{z,3}^E=0$.)
Note that $\mu_{z}^E$ is a fundamental linear-response property of the 2D crystal, in the sense that it can be computed within the primitive cell of the flat configuration~\cite{PhysRevLett.127.216801} without ambiguities about the electrical or mechanical boundary conditions.
Indeed, unlike in the 3D case of Sec.~\ref{sec.flexoel_bulk}, the structure of a bent 2D crystal is well defined in the static limit (think of rolling up the layer into a tube), and there are no far-away surface terms that may influence its electrical state.

As in 3D, $\mu^{E}_z$ can be separated into an electronic (clamped-ion) and a lattice-mediated part: the former survives even in graphene and silicene where no ionic mechanism exists.
Interestingly, the electronic response contains a purely geometric ``metric'' term, given by the quadrupole moment of the unperturbed charge density.  %
Such a contribution is not small; on the contrary, it nearly cancels out with the dipole moment of the first-order charge density as calculated in the curvilinear frame of the bent crystal.
One can verify (see the Supplemental Material of Ref.~\cite{PhysRevLett.127.216801}) that the resulting definition of $\mu^{E}_z$ correctly yields a vanishing response in an inert system, e.g. a fictitious layer of noninteracting spherical atoms; accounting for the metric term is essential for this result to hold.
{In $D_{3d}$ crystals an in-plane response is also allowed,} and takes the following form~\cite{PhysRevLett.131.236203,PhysRevLett.133.199901}
\begin{eqnarray}\label{eq.Pip}
    \mathbf{P}_\mathrm{ip} = \mu_\mathrm{ip} \left[ \sin(3\theta)\mathbf{\lambda}_1 + \cos(3\theta)\mathbf{\lambda}_2\right] \left( K_1 - K_2 \right).
\end{eqnarray}
Here $K_i$ are the principal curvatures; $\mathbf{\lambda}_1 = [\cos(\theta) , \sin(\theta)]$ and $\mathbf{\lambda}_2 = [-\sin(\theta) , \cos(\theta)]$ the corresponding eigenvectors of the curvature tensor (see Sec.~\ref{sec.curv}); $\theta$ is the angle between $\lambda_1$ and the $x$ direction, normal to the $yz$ mirror plane.
The modulus of the polarization vector is independent of the bending direction and proportional to the difference in principal curvatures; thus, the response vanishes for a spherical curvature.
Note the characteristic three-fold $\theta$-dependence in Eq.~\ref{eq.Pip}, reflecting the three-fold axis of the crystal structure. 
Such directional dependence allows strain engineering of polarization textures with a nontrivial topology~\cite{PhysRevLett.131.236203}.

The relative weights of the electronic and lattice-mediated contributions vary strongly across the aforementioned material classes, in both magnitude and sign. 
In the out-of-plane case, the clamped-ion response generally dominates over the lattice-mediated one in the transition metal dichalcogenides, while the converse is true for BN; SnS$_2$ seems to lie in the middle~\cite{PhysRevLett.127.216801}.
The in-plane response appears systematically larger than the out-of-plane one, likely due to the significant anisotropy in the polarizability of the layer~\cite{PhysRevLett.131.236203}.

\subsection{First Principles Theory and Computational Developments }\label{sec.flexoel_theory}

As we anticipated in the introduction, achieving a viable theoretical and computational description of flexoelectricity from first principles required addressing and solving 
many conceptual and practical difficulties.
In this Section we provide a brief overview of the main milestones that marked this journey, with an emphasis on the more recent developments. A more detailed account of the early breakthroughs can be found in Ref.~\cite{Stengel_Vanderbilt}.

The primary and most daunting challenge facing a calculation of flexoelectricity for a realistic crystal is the breakdown of translational symmetry that is inherent to a strain gradient.
Early attempts tackled this issue by constructing supercell geometries with the local strain varying along one of the spatial dimensions~\cite{Hong_2010,PhysRevB.84.180101,PhysRevB.88.174107}. While in principle possible, supercell strategies are computationally expensive and formally unpalatable: any bulk material property should be defined and calculated on the primitive crystal cell~\cite{PhysRevLett.105.127601}.

A path towards the solution came by representing the deformation field in terms of its Fourier components, the acoustic phonon modes, and analyzing their associated polarization current in the long-wavelength limit~\cite{PhysRevB.88.174106,PhysRevB.88.174107}. 
The practical advantage of the Fourier representation lies in a powerful characteristic of density-functional perturbation theory (DFPT)~\cite{Baroni2001}, 
where the linear response to an arbitrary modulated perturbation can be calculated on the primitive cell of the crystal: the incommensurate phases are conveniently reabsorbed as shifts in momentum space.

To capture flexoelectricity, however, a number of additional methodological advances were necessary. First, the use of coordinate transformations to the co-moving frame of the acoustic phonon~\cite{Stengel2013NatCommun,PhysRevB.98.125133,PhysRevB.99.085107} was key to understanding the role of surface piezoelectricity and of rototranslational symmetries. 
Next, the formulation of the macroscopic polarization response in an inhomogeneous context required a careful treatment of non-local pseudopotentials in the current-density operator~\cite{PhysRevB.98.075153}.
(Some issues remained with the breakdown of Galilean invariance and Larmor theorem, also related to non-local pseudopotentials, and were addressed only recently~\cite{Stengel-26}.)
Finally, a long-wavelength perturbative scheme was developed on top of DFPT~\cite{PhysRevX.9.021050,PhysRevB.105.064101} by promoting the wave vector of the perturbation to an additional small parameter. 
The combination of all the above advances, which constitutes the basis of the current ABINIT~\cite{abinit2020,abinit2025} implementation, allowed the flexoelectric tensor components to be computed all at once for an arbitrary crystal structure via an automated procedure.

We note that most of the above advances in first-principles theory and methods are far more general than the specifics of the flexoelectric problem: they allow for the calculation of the second derivative of the energy with respect to two arbitrary perturbations 
at first order in the wave vector ${\bf q}$. As a consequence, they led to significant progress in other areas of research as well.
Notable examples include the calculation of dynamical quadrupoles~\cite{PhysRevX.9.021050}, phonon magnetic moments~\cite{PhysRevB.105.094305} and 
natural optical activity~\cite{PhysRevLett.131.086902}. 
We are currently exploring their generalization to flexomagnetism (discussed in the next Section) %
as well, by combining the long-wavelength approach with some recent advances in the treatment of coupled spin-lattice dynamics~\cite{Royo-26}.

\section{Flexomagnetism} \label{sec.flexomag}

In a strict sense, the flexomagnetic effect is defined  
as the linear magnetization response to a strain gradient,
\begin{equation}\label{eq.flexomag_tensor}
    \mu_{\alpha\lambda,\beta\gamma}^M = \frac{\partial M_\alpha}{\partial \varepsilon_{\beta\gamma,\lambda}}.
\end{equation}
The direct analogy to Eq.~\ref{eq.flexoel_tensor} is clear.
Unlike flexoelectricity, however, this 
is far from being a universal effect. Indeed, it requires a magnetic point group where
both time-reversal and space-inversion symmetry are simultaneously broken: all linear magnetoelectrics (Sec.~\ref{sec.ME}) belong to this category. 
Perhaps for this reason, flexomagnetism has received much less attention than its electrical counterpart. Only recently was it indirectly detected in films of the linear magnetoelectric material Cr$_2$O$_3$, where strain gradients naturally occur by epitaxial strain relaxation~\cite{Makushko2022}. 
To date, we are not aware of any direct measurements of the 
flexomagnetic coefficients as defined in Eq.~\eqref{eq.flexomag_tensor}. 
We shall refer to this coupling as the \emph{linear flexomagnetic effect} henceforth, and discuss it more extensively 
in Sec.~\ref{sec.lin_flexomag}.

Some authors use the term flexomagnetism to describe a much broader range of phenomena where the spin Hamiltonian is modified by strain gradients or curvature, 
whether or not it results in a net magnetization~\cite{TANG2025100878}. 
This less restrictive definition includes other effects that are, unlike the linear flexomagnetic response of Eq.~\eqref{eq.flexomag_tensor}, universal in magnetic materials.
An important example is the Dzyaloshinskii-Moriya interaction (DMI) induced at linear order in the strain gradient, which is allowed by symmetry in any magnet.
This enables the geometric control of a rich variety of phenomena, including magnetochirality~\cite{hertel}.
Experimentally, a strain gradient of order $10^5~\mathrm{m}^{-1}$ in 50--200~nm La$_{0.67}$Sr$_{0.33}$MnO$_3$ films epitaxially grown on NdGaO$_3$~\cite{PhysRevLett.127.117204} was shown to induce
a DMI, detected by non-reciprocal spin-wave propagation, large enough to stabilize a room-temperature skyrmion lattice.
The counterpart of this effect in 2D magnets, the curvature-induced DMI, is the topic of Sec.~\ref{sec.intrinsicDMI}.

Before delving into the two manifestations of flexomagnetism that we have just introduced, we briefly review a number of essential background concepts.
These include an introduction to the basic models of magnetism (Sec.~\ref{sec.mag_bg}), and an overview of the magnetoelectric effect (Sec.~\ref{sec.ME}). 
The latter, in a way, bridges flexoelectricity and flexomagnetism, since strain gradients act on materials as effective electric fields. 
We conclude this section with a short mention of other phenomena that fall under the umbrella of flexomagnetism in the broader sense specified above
(Sec.~\ref{sec.otherFxM}).

\subsection{Background}\label{sec.mag_bg}

\subsubsection{Atomistic Models of Magnetism:} \label{sec.magmodels_atom}
A general anisotropic, bilinear, atomistic spin Hamiltonian is~\cite{RevModPhys.95.035004}
\begin{eqnarray}\label{eq.atomistic_mag}
    H & = & -\frac{1}{2}\sum_{i\neq j}S_i^{\alpha}\mathcal{J}_{ij}^{\alpha\beta}S_j^\beta - \sum_i S_i^{\alpha}\mathcal{A}^{\alpha\beta}_iS_i^{\beta}
\end{eqnarray}
for interaction between spins $\mathbf{S}_i$ at sites $i$, where the first term is interatomic, anisotropic exchange interaction $\mathcal{J}_{ij}^{\alpha\beta}$, and the second term is single-ion anisotropy $\mathcal{A}^{\alpha\beta}_i$. The tensor $\mathcal{J}_{ij}^{\alpha\beta}$ can be decomposed into isotropic Heisenberg exchange $J_{ij} = \frac{1}{3} \mathrm{Tr} \mathcal{J}_{ij}$ and symmetric (traceless) and antisymmetric parts. 
The antisymmetric part defines the Dzyaloshinskii–Moriya vector $D_{ij}^\alpha = -\frac{1}{2} \varepsilon_{\alpha\beta\gamma} \mathcal{J}_{ij}^{\beta\gamma}$, commonly described by the Hamiltonian term 
\begin{equation}\label{eq.atomicDMI}
    H_{D} = \sum_{i<j} \mathbf{D}_{ij} \cdot (\mathbf{S}_i \times \mathbf{S}_j). 
\end{equation}
 The DMI was introduced by Dzyaloshinskii~\cite{DZYALOSHINSKY1958241} to explain weak ferromagnetism in certain antiferromagnetic materials, such as $\mathrm{\alpha}$-Fe$_2$O$_3$.  Moriya explained the microscopic mechanism as a spin-orbit coupling (SOC) mediated super-exchange interaction~\cite{PhysRev.120.91}. The DMI vector $\mathbf{D}_{ij}$ is antisymmetric in $i \leftrightarrow j$ and its direction is determined by the Moriya rules~\cite{PhysRev.120.91}, one of which stipulates that the vector vanishes when there is an inversion center at the bond between sites $i$ and $j$.

Beyond weak ferromagnetism, the DMI stabilizes spin spirals~\cite{Dzyaloshinskii1964,P_Bak_1980}, renders spin-wave dispersions asymmetric~\cite{PhysRevLett.104.137203}, and underpins topological textures such as skyrmions~\cite{BogdanovYablonskii1989,Skyrmion_Science2009,Yu2010,GOBEL20211} and hopfions~\cite{10.1063/5.0099942,Zheng2023}, with prospective applications ranging from racetrack memories~\cite{Fert2013} to neuromorphic and reservoir computing~\cite{Song2020,Beneke2024}. Any perturbation that breaks inversion symmetry offers a handle on the DMI, and hence on these states; as we shall see, curvature and strain gradients offer a particularly versatile one.

\subsubsection{Continuum Models of Magnetism:} \label{sec.magmodels_cont}
We characterize magnetically ordered materials with magnetic order parameters. For a two-sublattice system we write the magnetization $\mathbf{M} = \frac{1}{2V} \left( \mathbf{m}_1 + \mathbf{m}_2 \right)$ or N\'{e}el vector $\mathbf{L} = \frac{1}{2V} \left( \mathbf{m}_1 - \mathbf{m}_2 \right)$ as the ferromagnetic (FM) and antiferromagnetic (AFM) order parameters, respectively, where $\mathbf{m}_i = -\frac{g_i \mu_\mathrm{B}}{\hbar} \mathbf{S}_i$ is the magnetic moment at site $i$ and $V$ is the volume per atom. 
In the continuum limit, where the order parameter varies slowly relative to interatomic distances, the energy in Eq.~\ref{eq.atomistic_mag} takes the form
\begin{eqnarray}\label{eq.cont_mag}
   E = \int d^3 r \left[ \mathcal{A}_{\alpha\beta\gamma\delta} \partial_\alpha M_\gamma \partial_\beta M_\delta + \mathcal{K}_{\alpha\beta} M_\alpha M_\beta  \right. \nonumber \\ \left. + \mathcal{D}_{\alpha\beta\gamma} \left( M_\alpha \frac{\partial M_\beta}{\partial r_\gamma} - M_\beta \frac{\partial M_\alpha}{\partial r_\gamma} \right)  \right] , 
\end{eqnarray}
for an order parameter $\mathbf{M}$. %
The first term is the exchange energy, which takes the common form $A \left[ \nabla \mathbf{M} \right]^2$ in isotropic media. %
The second term is the magnetic anisotropy, which  %
takes the form $K M_z^2$ in uniaxial magnets (with at least three-fold rotation symmetry around the axis $\hat{z}$).  The final term is a Lifshitz invariant whose coefficient
$\mathcal{D}_{\alpha\beta\gamma}=-\mathcal{D}_{\beta\alpha\gamma}$ is a
continuum counterpart of the DMI, as discussed below.

Each of the tensors in Eq.~\ref{eq.cont_mag} can be expressed in terms of the
parameters of Eq.~\ref{eq.atomistic_mag}~\cite{PhysRevB.100.214406}: the
exchange stiffness $A$ derives from the isotropic part of
$\mathcal{J}_{ij}$, and the anisotropy $\mathcal{K}_{\alpha\beta}$ from the
single-ion anisotropy and the symmetric anisotropic part of
$\mathcal{J}_{ij}$. For a Bravais lattice of spins, the coefficient
$\mathcal{D}_{\alpha\beta\gamma}$, which is also related to the
spiralization~\cite{Freimuth_2014}, is the first moment of the atomistic DMI
vectors, $\sum_j \mathbf{D}_{ij}\,r^\gamma_{ij}$; in the remainder of this
chapter we refer to it as the continuum DMI. In Sec.~\ref{sec.curvlinmag} we
will see that a term of the same form can also emerge from a curvilinear
coordinate transformation of the isotropic exchange alone. %

In an AFM material, the energetics of the primary order parameter can be expressed in the same form as Eq.~\ref{eq.cont_mag}, only replacing $\mathbf{M}$ with $\mathbf{L}$. 
This implies that many characteristics of ferromagnets can be translated to the AFM case directly, e.g., the possible emergence of topological noncollinear structures in the presence of a macroscopic DMI.
Nevertheless, it is the coexistence and mutual coupling of the order parameters $\mathbf{M}$ and $\mathbf{L}$ that provide the richest playground for interesting physics.
The simplest bilinear coupling, $\lambda_{\alpha\beta} L_\alpha M_\beta$, is precisely the term that Dzyaloshinskii introduced to explain weak ferromagnetism~\cite{DZYALOSHINSKY1958241}. 
(In a two-sublattice AFM crystal, the antisymmetric part of $\lambda_{\alpha\beta}$ originates from the microscopic inter-sublattice DMI, 
 while its symmetric
part stems from the sublattice asymmetry 
in single-ion anisotropy and symmetric anisotropic exchange.) 
Expanding $\lambda_{\alpha\beta}$ to leading order in perturbations such as strain $\varepsilon_{\gamma\delta}$, electric field $E_\gamma$ or curvature $s_n$, yields
\begin{equation}\label{eq.lambda_expand}
    \lambda_{\alpha\beta} (\varepsilon_{\gamma\delta} , E_\gamma , s_n)= \lambda_{\alpha\beta}^{0} +  \lambda_{\alpha\beta\gamma\delta}^\mathrm{piezo} \varepsilon_{\gamma\delta} + \lambda_{\alpha\beta\gamma}^\mathrm{ME} E_{\gamma} + \lambda_{\alpha\beta n}^\mathrm{FxM} s_n  + ...
\end{equation}
This relation describes how weak ferromagnetism can be induced linearly by a variety of external fields: piezomagnetism~\cite{Dzyaloshinskii1958piezomagnetism} responds to strain, the linear
ME effect (Sec.~\ref{sec.ME}) to an electric field, and the linear flexomagnetic effect (Sec.~\ref{sec.lin_flexomag}) to curvature.

\subsubsection{Magnetoelectric Effect:}
\label{sec.ME}

Here we briefly describe the linear magnetoelectric effect, i.e. the bilinear coupling between electric ($E_\beta$) and magnetic ($H_\beta$) fields.  %
It is typically described by expanding the free energy as follows~\cite{Fiebig_2005},%
\begin{equation}
    F(\mathbf{E},\mathbf{H}) = F_0 - \frac{1}{2}\chi_{\beta\gamma}^E E_\beta E_\gamma - \frac{1}{2}\chi_{\beta\gamma}^M H_\beta H_\gamma - \alpha_{\beta\gamma} E_\beta H_\gamma,
\end{equation}
where $\chi_{\beta\gamma}^E$ and $\chi_{\beta\gamma}^M$ are the electric and magnetic susceptibilities and $\alpha_{\beta\gamma}$ is the magnetoelectric tensor. 
Differentiation gives
\begin{align}
    P_\beta & = -\frac{\partial F}{\partial E_\beta} = \chi_{\beta\gamma}^E E_\gamma + \alpha_{\beta\gamma} H_\gamma   \label{eq.ME_P} \\
    M_\beta & = -\frac{\partial F}{\partial H_\beta} = \chi_{\beta\gamma}^M H_\gamma + \alpha_{\gamma\beta} E_\gamma.  \label{eq.ME_M}
\end{align}
Here, the terms containing $\alpha_{\beta\gamma}$ describe the linear magnetoelectric effect through which a magnetization $M_\beta$ is linearly induced by an electric field, and an electric polarization $P_\beta$ is linearly induced by a magnetic field. 
We can interpret the origin of this term from the expansion of $\lambda_{\alpha\beta}(E_\gamma)$ in Eq.~\ref{eq.lambda_expand}, as being mediated by the N\'{e}el vector: minimizing the free energy with respect to $\mathbf{M}$ and keeping only terms linear in $E$ gives $\alpha_{\gamma\delta} = -\,\chi^{M}_{\delta\beta}\, \lambda^\mathrm{ME}_{\alpha\beta\gamma}\, L^{0}_\alpha$, where $\mathbf{L}^{0}$ is the equilibrium N\'{e}el vector in the absence of
external fields. Its magnitude, set by the exchange interactions and the temperature, enters $\alpha$ as a prefactor, so that $\alpha$ vanishes above $T_\mathrm{N}$; the field-induced change of $\mathbf{L}$ is of
second order in $E$ and does not contribute to the linear response.

The magnetoelectric effect is described by a rank-two axial tensor, $\alpha_{\beta\gamma}$, and can only be non-zero in magnetic point groups that break both space inversion ($\mathcal{I}$) and time reversal ($\mathcal{T}$) symmetry. (Their product $\mathcal{I}\mathcal{T}$ need not be broken, though.)  Of the 69 magnetic point groups that contain neither $\mathcal{I}$ nor $\mathcal{T}$ (see Table~\ref{table.flexomagPG}), 58 allow a linear ME effect~\cite{Rivera2009}. 
The effect was predicted by Dzyaloshinskii~\cite{ME_Dzyaloshinsky} and first measured by Astrov~\cite{ME_Astrov1960,ME_Astrov} in Cr$_2$O$_3$, whose magnetic space group $R\bar{3}'c'$ below $T_\mathrm{N}\approx 307$~K~\cite{VOLGER1952} allows two independent coefficients, $\alpha_{11}=\alpha_{22}=\alpha_\perp$ and $\alpha_{33}=\alpha_\parallel$.

Microscopically, the ME response separates into electronic (clamped-ion) and lattice-mediated (ionic) contributions, in analogy with flexoelectricity (Sec.~\ref{sec.flexoel_mech}), and into spin and orbital contributions to the moments~\cite{PhysRevLett.101.117201,PhysRevB.86.094430}. Both channels reach the spins through the field-induced changes of the single-ion anisotropy and $g$-tensor~\cite{PhysRevLett.6.609}, of the Dzyaloshinskii-Moriya exchange, and of the isotropic exchange~\cite{Mostovoy2010}. In Cr$_2$O$_3$ the transverse response $\alpha_\perp$ is dominated by the relativistic channels, with electronic and ionic parts of comparable size and partly opposite sign~\cite{PhysRevLett.101.117201,PhysRevB.86.094430,Mu2014}, whereas $\alpha_\parallel$ vanishes at $T=0$ up to a small orbital term and arises at finite temperature from the field-induced modification of the isotropic exchange acting on thermally disordered spins, which produces its characteristic peak near $T_\mathrm{N}$~\cite{Mostovoy2010,Mu2014}.

Since flexoelectricity makes a strain gradient act on a crystal as an effective electric field, every linear ME material must also exhibit a magnetization linear in the strain gradient. The linear flexomagnetic effect of this form is the subject of Sec.~\ref{sec.lin_flexomag}.

\subsection{Linear Magnetization Response to Curvature}\label{sec.lin_flexomag}

\subsubsection{Definition and Motivation: }

In analogy to the electromechanical couplings in Eq.~\ref{eq.flexoel}, we can write the leading-order magnetomechanical couplings as
\begin{equation}\label{eq.linflexmag}
M_\alpha = q_{\alpha\beta\gamma}\,\varepsilon_{\beta\gamma} + \mu^{M}_{\alpha\beta\gamma\delta}\,\frac{\partial \varepsilon_{\gamma\delta}}{\partial r_\beta},
\end{equation}
where the first term describes piezomagnetism and the second is the linear flexomagnetic effect.
Both offer a route to controlling magnetization by purely mechanical means, without magnetic fields or currents. 
Yet, as discussed in Sec.~\ref{sec.fxM_sym}, the two tensors are subject to symmetry constraints quite unlike those of their electric counterparts.
As we saw in the section on flexoelectricity, couplings to strain gradients are difficult to define unambiguously as 3D bulk properties: we expect surface piezomagnetism, dynamic effects, as well as the lack of a well-defined electrostatic reference potential to affect the magnetic case in a similar way. For bendable 2D membranes, which do not pose such problems, we write the magnetic analogue of Eq.~\ref{eq.flexoel_2D}
\begin{equation}\label{eq.flexomag_2D}
M_\alpha = \mu_{\alpha, n}^M s_n .
\end{equation}

\subsubsection{Symmetry Properties: }\label{sec.fxM_sym} 

In Eq.~\ref{eq.linflexmag}, $q_{\alpha\beta\gamma}$ and $\mu_{\alpha\beta\gamma\delta}^M$ are rank three and rank four axial tensors, odd under time-reversal and symmetric in the last two indices.
Their symmetry properties are therefore distinctly different from the piezoelectric and flexoelectric tensors in Eq.~\ref{eq.flexoel}, which are invariant under time reversal symmetry 
(see Table~\ref{table.sym}). 
Importantly, $\mu_{\alpha\beta\gamma\delta}^M$
is odd under time reversal 
and, being an axial tensor of even order, it is also forbidden by space inversion. 
All in all, linear flexomagnetism is only allowed in the 69 magnetic point groups listed in Table~\ref{table.flexomagPG}~\cite{PhysRevB.84.174112,mtensor}. %

The universality of flexoelectricity ensures that a strain gradient always induces a polarization. In the ME materials, this polarization is accompanied by a magnetization. Consequently,  
there is a linear flexomagnetic response in all 58 ME point groups, as is also reflected in Table~\ref{table.sym}, where the flexomagnetic and ME tensors share symmetry behaviour under time and space inversion. %
This leaves 11 additional groups, which are marked in bold in Table~\ref{table.flexomagPG}, that allow a flexomagnetic response, but not a linear ME effect. In these 11 flexomagnetic, but non-ME point groups, other symmetry operations than time and space inversion enforce all elements of a rank 2 tensor to be zero, while additional degrees of freedom in a rank 4 tensor allow non-zero elements. These are relatively high-symmetry groups, found among cubic and hexagonal crystal systems.
The piezomagnetic effect differs in that it is even under space inversion but odd under the product of time and space inversion symmetry. Hence, some ME/flexomagnetic point groups that do not contain $IT$ allow also for piezomagnetism, while those piezomagnetic point groups that contain $I$ enforce the flexomagnetic/ME tensors to be zero. 

\begin{table}[ht]
\centering
\caption{Magnetic point groups that allow non-zero linear flexomagnetic tensor $\mu^{M}_{\alpha\beta\gamma\delta}$.
Groups in $\mathbf{bold}$ do not allow the linear magnetoelectric (ME) tensor $\alpha_{\beta\gamma}$.}
\label{table.flexomagPG}
\begin{tabular}{|l|p{9.2cm}|}
\hline
\textbf{Crystal System} & \textbf{Flexomagnetic point groups } \\
\hline\hline

Triclinic
& \(\bar{1}^{\prime}\), \(1\) \\
\hline

Monoclinic
& \(2^{\prime}\), \(2^{\prime}/m\), \(2\), \(2/m^{\prime}\), \(m^{\prime}\), \(m\) \\
\hline

Orthorhombic
& \(2^{\prime}2^{\prime}2\), \(222\), \(m^{\prime}m^{\prime}2\), \(m^{\prime}m^{\prime}m^{\prime}\),
  \(m^{\prime}m2^{\prime}\), \(m^{\prime}mm\), \(mm2\) \\
\hline

Tetragonal
& \(\bar{4}^{\prime}\), \(\bar{4}^{\prime}2^{\prime}m\), \(\bar{4}^{\prime}2m^{\prime}\), \(\bar{4}\),
  \(\bar{4}2^{\prime}m^{\prime}\), \(\bar{4}2m\),
  \(4^{\prime}\), \(4^{\prime}/m^{\prime}\), \(4^{\prime}/m^{\prime}m^{\prime}m\),
  \(4^{\prime}22^{\prime}\), \(4^{\prime}m^{\prime}m\),
  \(4\), \(4/m^{\prime}\), \(4/m^{\prime}m^{\prime}m^{\prime}\),
  \(4/m^{\prime}mm\), \(42^{\prime}2^{\prime}\), \(422\), \(4m^{\prime}m^{\prime}\), \(4mm\) \\
\hline

Trigonal
& \(\bar{3}^{\prime}\), \(\bar{3}^{\prime}m\), \(\bar{3}^{\prime}m^{\prime}\),
  \(3\), \(32^{\prime}\), \(32\), \(3m^{\prime}\), \(3m\) \\
\hline

Hexagonal
& \(\bar{6}^{\prime}\), \(\bar{6}^{\prime}m^{\prime}2\), \(\bar{6}^{\prime}m2^{\prime}\),
  \(\mathbf{\bar{6}}\), \(\mathbf{\bar{6}m^{\prime}2^{\prime}}\), \(\mathbf{\bar{6}m2}\),
  \(\mathbf{6^{\prime}}\), \(\mathbf{6^{\prime}/m}\), \(\mathbf{6^{\prime}/mmm^{\prime}}\),
  \(\mathbf{6^{\prime}22^{\prime}}\), \(\mathbf{6^{\prime}mm^{\prime}}\),
  \(6\), \(6/m^{\prime}\), \(6/m^{\prime}m^{\prime}m^{\prime}\),
  \(6/m^{\prime}mm\), \(62^{\prime}2^{\prime}\), \(622\), \(6m^{\prime}m^{\prime}\), \(6mm\) \\
\hline

Cubic
& \(\bar{4}^{\prime}3m^{\prime}\), \(\mathbf{\bar{4}3m}\), \(23\), \(\mathbf{4^{\prime}3 2^{\prime}}\), \(432\),
  \(m^{\prime}\bar{3}^{\prime}\), \(\mathbf{m^{\prime}\bar{3}^{\prime}m}\), \(m^{\prime}\bar{3}^{\prime}m^{\prime}\) \\
\hline

\end{tabular}
\end{table}

Lukashev \emph{et al.}~\cite{PhysRevB.82.094417} presented one of the first
studies of the linear flexomagnetic effect, using first-principles simulation methods
on a supercell of Mn$_3$GaN (space group $Pm\bar{3}m$), which orders in a noncollinear structure of magnetic point group $\bar{3}m$ below 298~K~\cite{doi:10.1143/JPSJ.44.781}. %
The calculated magnetic moment per unit strain gradient (of about $2~\mu_\mathrm{B}$\AA{} 
per unit cell) is, however, unlikely to reflect a bulk property of the material: 
$\bar{3}m$ breaks time reversal but retains inversion, so it admits a piezomagnetic response while the bulk linear flexomagnetic tensor vanishes by
symmetry~\cite{mtensor}. 
This mirrors the difficulty, discussed above for
flexoelectricity, of defining and computing flexo-effects as static bulk
properties of three-dimensional crystals, and motivates turning to
two-dimensional materials. Since the discovery of
long-range magnetic order in atomically thin van der Waals
crystals~\cite{Lee2016,Huang2017,CGT_gong}, a large family of 2D
ferromagnets and antiferromagnets has become available~\cite{Burch2018,Gibertini2019}, and recent high-throughput
calculations of magnetoelectric monolayers~\cite{Mangeri_2026} provide a
natural pool of candidates for the linear flexomagnetic effect.

Considering the bending of a 2D layer, the cubic point groups have no 2D counterpart, leaving 61 magnetic point groups as candidates for 2D flexomagnetism. We saw in Sec.~\ref{sec.curv} that the curvatures $s_n$ correspond to components $\varepsilon_{\alpha\beta,z}$, and the flexomagnetic coupling $\mu_{\alpha, n}^M$ in Eq.~\ref{eq.flexomag_2D} is correspondingly a sub-block of that in Eq.~\ref{eq.linflexmag}. Its symmetry constraints are therefore stricter: of the 61 layer-compatible groups in Table~\ref{table.flexomagPG}, eight ($4mm$, $42'2'$, $4/m'mm$, $\bar{4}'2'm$, $6mm$, $62'2'$, $6/m'mm$ and $\bar{6}'m2'$) allow a flexomagnetic response to in-plane strain gradients but none to curvature. %

\subsubsection{Mechanisms: }

In Eq.~\ref{eq.lambda_expand}, we saw the flexomagnetic effect emerge as a curvature-induced coupling between AFM and FM order parameters, through a term 
\begin{equation}\label{eq.F_fxm}
F = \lambda_{\alpha\beta n}^\mathrm{FxM}  L_\alpha M_\beta s_n
\end{equation}
in the energy density, revealing ${\mu}^{M}_{\alpha, n} = -\,\chi^{M}_{\alpha\delta}\,\lambda^{\mathrm{FxM}}_{\beta\delta n}\, L_\beta$ as mediated by the symmetry breaking AFM order. Since $\mathbf{M}$ is even and $s_n$ odd under space inversion, Eq.~\ref{eq.F_fxm}
is only invariant if $\mathbf{L}$ is odd under inversion, i.e. if inversion
exchanges the two AFM sublattices, consistent with the absence of $I$ in all
groups of Table~\ref{table.flexomagPG}. Combining Eq.~\ref{eq.flexoel_2D} and Eqs.~\ref{eq.ME_P}-\ref{eq.ME_M} implies a magnetization contribution   
\begin{equation}
M_\alpha = {\alpha}_{\beta\alpha}\,\left({\chi}^{E}\right)^{-1}_{\beta\gamma} {\mu}^{E}_{\gamma,n}  s_n, 
\label{eq.FxM_FxE_ME}
\end{equation}
which can be seen as one possible mechanism for the linear flexomagnetic effect, where polar distortions induced by curvature induce magnetization through the ME effect. It is clear that there must also be direct contributions to ${\mu}^{M}_{\alpha, n}$, that cannot be mediated by magnetoelectricity and which make up the full effect in the eight layer-compatible groups of Table~\ref{table.flexomagPG} that allow a non-zero flexomagnetic tensor but enforce a zero ME tensor, i.e. the eight hexagonal groups in bold.

The mechanisms of flexomagnetism are expected to combine features of those discussed for flexoelectricity and for the ME effect. As in flexoelectricity (Sec.~\ref{sec.flexoel_mech}), the strain gradient acts through two channels: an ionic channel, in which the forces on the polar lattice modes produce internal sublattice displacements, and an electronic (clamped-ion) channel, in which the inhomogeneous deformation modifies the electronic structure at fixed ionic positions. As in the ME effect (Sec.~\ref{sec.ME}), each channel reaches the magnetization through the gradient-induced changes of the atomistic magnetic couplings in Sec.~\ref{sec.magmodels_atom} %
or through a change of the $g$-tensor, i.e.\ of the relation $\mathbf{m}_i = -(g_i\mu_\mathrm{B}/\hbar)\mathbf{S}_i$ between moment and spin. The latter, orbital channel produces a net magnetization from a perfectly compensated spin structure if the change is sublattice-odd. This was experimentally estimated to contribute $10~\mu_\mathrm{B} \, \mathrm{nm}^{-2}$ to the flexomagnetic coefficient of Cr$_2$O$_3$ films, where the dominating contribution of $15~\mu_\mathrm{B} \, \mathrm{nm}^{-2}$ was suggested to originate from the thermodynamic effect of a graded N\'{e}el temperature, whereby the sublattice magnetizations are no longer compensated at a given temperature~\cite{Makushko2022}. The latter results from the strain dependence of the isotropic exchange interaction. 

Since $\lambda^{\mathrm{FxM}}_{\alpha\beta n}$ was introduced in Eq.~\ref{eq.lambda_expand} as the curvature dependence of the coupling $\lambda_{\alpha\beta}$ between $L_\alpha$ and $M_\beta$, its microscopic content %
follows from that of $\lambda_{\alpha\beta}$. %
The components of $\lambda^{\mathrm{FxM}}_{\alpha\beta n}$ can therefore be expressed in terms of the curvature derivatives of $\mathcal{A}^{\alpha\beta}_i$ and the anisotropic parts of $\mathcal{J}^{\alpha\beta}_{ij}$ in Eq.~\ref{eq.atomistic_mag}. A curvature-induced magnetization thus requires bending to render the two sublattices inequivalent at the level of these parameters, the atomistic counterpart of the requirement, following from Eq.~\ref{eq.F_fxm}, that $\mathbf{L}$ be odd under the inversion operation that curvature removes.

Returning to our example of 2D materials of $D_{3d}$ symmetry, the coupling in Eq.~\ref{eq.F_fxm} is symmetry allowed in these crystals when $\mathbf{L}$ is inversion odd. In addition to the prototypical ferromagnetic monolayer CrI$_3$, the $D_{3d}$ crystal class includes a large group of AFMs~\cite{Burch2018,PhysRevLett.124.027601,Olsen_2021,10.1063/5.0206361,Ni2021_MnPSe3,Olsen_2024}. Those with N\'{e}el type AFM order, such as MnP(S/Se)$_3$, are linear ME materials~\cite{10.1063/5.0206361,PhysRevLett.124.027601}, and consequently support the linear flexomagnetic effect. This example, for out-of-plane magnetic moments, is shown in Fig.~\ref{fig.ME_flexo_combo}(a)-(c), which also illustrates the symmetry relation between (a) flexoelectricity: bending around the $x$-axis induces an in-plane electric polarization along $\hat{x}$, in addition to the out-of-plane polarization (unlike in Eq.~\ref{eq.Pip}, where $\hat{x}$ is normal to a mirror plane, we now align $\hat{x}$ with the zigzag direction, following Ref.~\cite{PhysRevLett.128.177202}), (b) the linear ME effect couples $P_x$ to magnetization component $M_y$, and (c) the linear flexomagnetic effect: bending around $\hat{x}$ induces magnetization along $\hat{y}$ consistent with (a) and (b) combined. %

\begin{figure}[h!]
	\centering
	\includegraphics[width=0.98\textwidth]{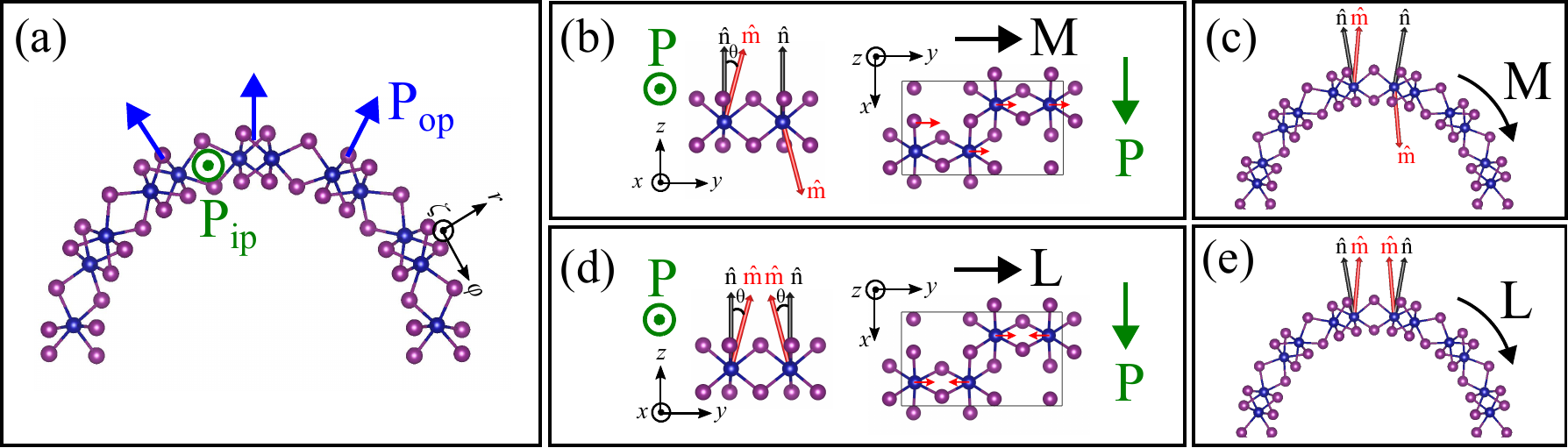}
	\caption{ Illustration of ME and flexo-effects in a $D_{3d}$ monolayer (a) Flexoelectricity induces both out-of-plane and in-plane components of polarization as a response to bending. (b) Linear ME coupling between electric polarization $\mathbf{P}$ and magnetization in AFM. (c) A linear flexomagnetic coupling between curvature and magnetization in an AFM. (d)-(e) show the corresponding effects in a FM, where a N\'{e}el vector couples to (d) electric polarization or (e) curvature. 
    The curvilinear coordinates $(\zeta, \varphi, r)$ of the cylindrically bent surfaces locally correspond to Cartesian coordinates $(x,y,z)$ in the flat monolayer. 
    }
	\label{fig.ME_flexo_combo}
\end{figure}

Since Eq.~\ref{eq.F_fxm} contains $\mathbf{L}$ and $\mathbf{M}$ on an equal footing, it permits an analogous counterpart of the linear flexomagnetic effect in ferromagnets, manifesting itself as a N\'{e}el vector induced linearly with curvature. This effect is expected to occur in e.g. CrI$_3$, where the ME analogue [see Fig.~\ref{fig.ME_flexo_combo}(d)] was recently discussed in Ref.~\cite{Edstrom2025}, and the flexomagnetic counterpart [Fig.~\ref{fig.ME_flexo_combo}(e)] can in principle be deduced from DFT calculations on curved nanotube geometries such as those in Ref.~\cite{PhysRevLett.128.177202}.

\subsection{Curvature-Induced DMI}\label{sec.intrinsicDMI}

\subsubsection{Definition and Motivation: }

We now turn to the curvature‑induced DMI in centrosymmetric crystals, where the continuum DMI term vanishes in undeformed configurations and under uniform strain. In acentric materials, the leading‑order effect on the DMI instead is its modification due to strain~\cite{PhysRevB.98.214414} (not gradients of strain), as considered in e.g. Ref.~\cite{PhysRevLett.128.257201}. %

{%
As mentioned above, 
strain gradients can be used to engineer DMI in bulk-like films as well~\cite{PhysRevLett.127.117204}, but in the following we again restrict ourselves to 2D layers, for which the flexo-coefficients are well-defined properties.
}
In the case of bending a 2D magnet, the tensor $\mathcal{D}_{\alpha\beta\gamma}$ linearly induced by curvature $s_n$ is
\begin{equation}
    \mathcal{D}_{\alpha\beta\gamma} = \Delta_{\alpha\beta\gamma,n} s_n,
\end{equation}
where $\Delta_{\alpha\beta\gamma,n}$ is antisymmetric in $\alpha \leftrightarrow \beta$ so $\Delta_{\alpha\beta\gamma,n}$ can have at most 18 independent components.

\subsubsection{Symmetry Properties: } \label{sec.DMI_sym}

The form of $\Delta_{\alpha\beta\gamma,n}$ in a crystal of a given symmetry is determined by Neumann's principle~\cite{birss1966symmetry}. In the highest symmetry layer groups (point group $D_{6h}$ of order 24), two independent components are non-zero: $D_1 = \Delta_{yzy,2} = \Delta_{xzx,1} $ and $D_2 =  \Delta_{xzx,2} = \Delta_{yzy,1}$ (the components involving $s_3$ are then fixed by in-plane isotropy). The former reduces to the ``geometric DMI'' (see next subsection) in the nonrelativistic limit,
and describes the linear-in-$q$ energy-dependence of a spin cycloid propagating azimuthally on the cylinder surface. 
The second component, $D_2$, has not been studied before in the curvilinear magnetism literature to the best of our knowledge. It is symmetry-allowed in all 2D crystals, and corresponds to the energy of a spin cycloid along the axis of a cylinder. Lower symmetry crystal classes allow yet additional terms, with $D_3 = \Delta_{xyy,2} = -\Delta_{xyy,1} = -\frac{1}{2} \Delta_{xyx,3}$ additionally being non-zero in the $D_{3d}$ class of 2D materials that we have discussed. Like $D_1$, $D_3$ is related to the energy of a spiral propagating around a cylinder, but with spins rotating in the plane.

\subsubsection{Mechanisms: }

\label{sec.curvlinmag}
   To illustrate the physical mechanisms that are responsible for
   the curvature-induced DMI, it is convenient to work in curvilinear coordinates,
   where the (quasi-2D) magnetic layer is treated as a surface embedded in 3D space
   and parametrized according to the guidelines of Sec.~\ref{sec.curv}.
   Recall that, in such a setting, the local coordinate frame varies in space and is
   defined at any given point ${\bf r}$ by the triad $\hat{\mathbf{u}}$, $\hat{\mathbf{v}}$, and $\hat{\mathbf{n}}$: these correspond, respectively,
   to the two tangential vectors and the normal direction $\hat{\mathbf{n}} = \hat{\mathbf{u}} \times \hat{\mathbf{v}} / \left| \hat{\mathbf{u}} \times \hat{\mathbf{v}} \right|$.
The magnetic energy density takes a similar form as Eq.~\ref{eq.cont_mag}
but while the magnetization is still a 3D vector constrained to the sphere, the position components $r_\alpha$ run over only the two in-plane coordinates, and the total energy is the 2D areal integral $E = \int w \mathrm{d}^2 r$, with areal energy density $w$. Importantly, the parameters $A$, $\mathcal{D}_{\alpha\beta\gamma}$, $\mathcal{K}_{\alpha\beta}$ and $\lambda_{\alpha\beta}$ become curvature-dependent quantities.   %

There have been significant developments in the field of curvilinear magnetism in the past decade, where a key theoretical foundation is the insight that applying a curvilinear coordinate transformation to the non-relativistic, isotropic exchange energy $w_\mathrm{ex} = A_0 \left[ \nabla \mathbf{M} \right]^2$ (first term of Eq.~\ref{eq.cont_mag}) leads to effective terms of the form of anisotropy and DMI~\cite{PhysRevLett.112.257203,Sheka_2015,curvilinmicromag2022}. 
Curvature therefore generally allows these terms to appear as geometric effects that depend only on the shape of the magnet, not on a deformation of its lattice. They circumvent the usual need for SOC or restrictions on crystallographic symmetries, but acquire physical consequences only when an anisotropy ties the magnetization to the local frame.
As a concrete example, we write the transformation of the exchange term in cylindrical coordinates $(\zeta, \varphi, r)$
\begin{align} \label{eq.exch_cyl}
    w_\mathrm{ex} & = A_0 \left(\left[ \partial_\varphi M_\alpha \right]^2 + \left[ \partial_\zeta M_\alpha \right]^2\right) + {\frac{{A_0}}{R^2} \left[ M_r^2 +  M_\varphi^2  \right]} +  \frac{2A_0}{R} \left[ M_r \partial_\varphi M_\varphi - M_\varphi \partial_\varphi M_r  \right] %
\end{align}
where $\partial_\varphi = \frac{1}{R}\frac{\partial}{\partial \varphi}$ and we see the effective anisotropy and DMI terms 
\begin{equation}\label{eq.noSOCpars}
{\mathcal{K}}={\mathcal{K}}_r={\mathcal{K}}_\varphi=\frac{A_0}{R^2} \qquad {\mathcal{D}}=2\frac{A_0}{R} ,  
\end{equation}
emerging %
from the curvilinear coordinate transformation. %
(In the above paragraphs we have based our discussion of the
geometric DMI on the continuum spin Hamiltonian; similar considerations
hold for the atomistic case as well.)

In a real magnet, of
   course, the purely geometric picture outlined above needs to be revised,
   and the relativistic effects of spin-orbit coupling properly accounted for when
   taking the perturbative expansion of the energy in powers of the local curvature.
   This is a calculation that can hardly be carried out at a phenomenological
   level, as the underlying physical ingredients (curvature-induced distortions of
   the bonding network, rearrangement of the charge and spin density following the
   deformation, etc.) are genuinely microscopic in nature. These effects have been computed 
 from first principles in CrI$_3$~\cite{PhysRevLett.128.177202}, with special attention to assessing the
   impact of SOC on curvature-induced DMI.
The strategy of Ref.~\cite{PhysRevLett.128.177202} relies on nanotube geometries
   of varying radius, where spin cycloids of commensurate periodicity are prepared
   along the azimuthal direction of armchair nanotubes. Calculations are carried
   out within noncollinear-spin DFT, either with or without SOC.

Turning off SOC yields ${\Delta}_{yzy,2} = 2A_0 = 99.9~\mathrm{meV\AA^2} / \mathrm{Cr}$, in agreement with the value expected for the geometric term according to Eq.~\ref{eq.noSOCpars}, with $A_0 = 49.9~\mathrm{meV\AA^2} / \mathrm{Cr}$. Calculations including SOC led to a value of ${\Delta}_{yzy,2} = 63.4~\mathrm{meV\AA^2} / \mathrm{Cr}$, reduced by more than a third. Thus, the intrinsic contribution to the curvature-induced DMI is of similar order of magnitude as the geometric term, and cannot be neglected. %
The other elements of $\Delta_{\alpha\beta\gamma,n}$, e.g. $D_2$ and $D_3$ above, have no geometric counterpart and are fully determined by the curvature dependence of intrinsic materials properties. Quantitatively determining their values requires materials-specific, first-principles calculations (e.g. DFT). However, while $D_1$ was calculated using supercells in Ref.~\cite{PhysRevLett.128.177202}, computing $D_2$ would require prohibitively expensive calculations of cycloids propagating along a nanotube. This illustrates the need for new method developments to calculate flexomagnetic coefficients from first principles, without the need for supercells.

Like the other effects discussed in this chapter, the intrinsic part of $\Delta_{\alpha\beta\gamma,n}$ separates into lattice-mediated and electronic (clamped-ion) contributions, and at the atomistic level it derives from the curvature derivatives of the DMI vectors $\mathbf{D}_{ij}$ of Eq.~\ref{eq.atomicDMI}, subject to the Moriya rules for the bent bond geometry. %

\subsection{Other Flexomagnetic Phenomena}\label{sec.otherFxM}

The broad definition of flexomagnetism as any coupling phenomenon between magnetism and strain gradients or curvature potentially embodies many more effects than those discussed so far. A few examples are briefly described below.

\subsubsection{Nonlinear Flexomagnetic Effects:}\label{sec.nonlin}

The effects of Secs.~\ref{sec.lin_flexomag} and \ref{sec.intrinsicDMI} are odd under inversion and therefore linear in curvature. Inversion-even quantities, such as the magnetic anisotropy, instead respond at second order in centrosymmetric crystals. Eq.~\ref{eq.noSOCpars} already provides a geometric, non-relativistic example; as for the DMI, curvature also modifies the intrinsic anisotropy by altering the atomic and electronic structure, and the breaking of crystallographic symmetries admits terms with no geometric counterpart. For a bent 2D crystal we write
\begin{equation}
    \mathcal{K}_{\alpha\beta}(s) - \mathcal{K}_{\alpha\beta}(0) = B_{\alpha\beta}^{nm} s_n s_m ,
\end{equation}
symmetric in $\alpha\beta$ and in $nm$, with at most 30 independent coefficients (one component of $\mathcal{K}_{\alpha\beta}$ is fixed by choosing a zero energy reference). At the atomistic level, the intrinsic part derives from the curvature dependence of the single-ion anisotropy $\mathcal{A}^{\alpha\beta}_i$ and of the symmetric anisotropic part of $\mathcal{J}^{\alpha\beta}_{ij}$.

For the $D_{3d}$ ferromagnet CrI$_3$, the diagonal components of $B_{\alpha\beta}^{22}$ were calculated in Ref.~\cite{PhysRevLett.128.177202} (there labeled $\mathcal{K}_{r,2} = B_{zz}^{22}-B_{xx}^{22}$ and $\mathcal{K}_{\varphi,2} = B_{yy}^{22} - B_{xx}^{22}$): without SOC they are equal and satisfy Eq.~\ref{eq.noSOCpars}, whereas with SOC they differ from each other and from Eq.~\ref{eq.noSOCpars} by roughly a factor of two. 
The $D_{3d}$ symmetry leaves six independent coefficients, of which two deserve comment. First, the off-diagonal coefficient $B^{11}_{xz}$ rotates the easy axis of magnetization upon bending, a second-order effect neglected in curvilinear magnetism that may become relevant when the radius of curvature approaches interatomic distances. Second, one of the six is the coupling of the uniaxial anisotropy to the Gaussian curvature, $K_1K_2 = s_1 s_2 - s_3^2$. Since every cylindrical deformation has $K_1K_2=0$, this coefficient is invisible to bending about any in-plane axis and can only be obtained from deformations with non-zero Gaussian curvature, in contrast to the DMI coefficients of Sec.~\ref{sec.DMI_sym}, which are all accessible from cylinders. This again underscores the need for methods that determine the full curvature dependence of the magnetic parameters without recourse to supercells.

\subsubsection{Inhomogeneous Magnetoelectric Effects: }\label{sec.inhom_ME}

Multiferroic materials simultaneously host a spontaneous electric polarization and magnetic order~\cite{Spaldin2019}.  %
In so-called type-II multiferroics~\cite{Khomskii2009}, the electric polarization arises due to symmetry breaking caused by the magnetic order. For example, certain spin spirals break inversion symmetry and induce an electric polarization~\cite{PhysRevLett.96.067601,PhysRevLett.107.157202,Tokura_spiralmulti} through an inhomogeneous magnetoelectric effect~\cite{inhom_magnetoel} 
\begin{equation}\label{eq.contP}
P_\alpha = {d}_{\alpha\beta,\gamma\delta}\left( M_\gamma \partial_\beta M_\delta - M_\delta \partial_\beta M_\gamma\right). 
\end{equation}
As can be anticipated from the resemblance to the continuum form of the DMI in Eq.~\ref{eq.cont_mag}, ${d}_{\alpha\beta,\gamma\delta}$ corresponds to the coefficient of the DMI that is linearly induced by an applied electric field~\cite{Edstrom2025}. 
Eq.~\ref{eq.contP} can be considered the electric-field-induced DMI analogue of the curvature-induced DMI discussed in Sec.~\ref{sec.intrinsicDMI}, in the same sense that the linear ME effect is analogous to the flexomagnetic effect of Eq.~\ref{eq.linflexmag}.

Eq.~\ref{eq.contP} is the antisymmetric part of a broader class of gradient couplings, whose symmetric counterpart includes flexoelectricity~\cite{PhysRevLett.132.146801}. Being a coupling of magnetism to a gradient, albeit not of strain, it has also been called ``spin flexoelectricity''~\cite{PYATAKOV2015255} or flexomagnetoelectricity~\cite{Pyatakov2009,TANYGIN2011616}. In this chapter we reserve flexomagnetism for the magnetomechanical coupling to strain gradients, and flexomagnetoelectricity for joint responses of polarization and magnetization to strain gradients, close to recent usage of the term to describe a strain-gradient-induced ME effect~\cite{PhysRevLett.133.156505}.

\subsubsection{Flexomagnetism without Magnetic Order:}\label{sec.rashba}
In non-centrosymmetric materials, SOC lifts the spin degeneracy of the electronic bands; the ensuing Rashba effect underlies a range of spin-transport phenomena central to spintronics~\cite{Manchon2015}. Since curvature breaks inversion symmetry, it can generate a Rashba splitting even in a centrosymmetric material: quantum-mechanical models of electrons confined to curved low-dimensional geometries~\cite{JENSEN1971586,PhysRevA.23.1982} yield a curvature-induced Rashba SOC~\cite{gentile1}, which in turn can drive topological electronic states and metal--insulator transitions in curved nanostructures~\cite{gentile2}.

By itself the Rashba splitting produces no magnetization, since time-reversal symmetry pairs opposite spin states at $\mathbf{k}$ and $-\mathbf{k}$ and makes the spin textures of the split bands cancel in equilibrium; a net spin polarization requires a driven state, as in the Edelstein effect, where a charge current through a Rashba conductor generates a magnetization~\cite{Edelstein1990}. A strain gradient can play the same role: acting as a mechanical driving force on the carriers of a non-centrosymmetric conductor, it induces a magnetization in close analogy with the Edelstein effect~\cite{koyama2026flexocurren}. Being a dissipative, nonequilibrium response, this ``flexocurrent" magnetization is allowed even in time-reversal-symmetric materials, provided the flexocurrent is continuously driven; in thermal equilibrium, time-reversal symmetry forbids it.
It should not be confused with the pseudomagnetic field of strained graphene~\cite{Guinea2010}, which is time-even, opposite in the two valleys, and induces no net magnetization.

\subsubsection{Altermagnetism and Related Magnetic Orders: } \label{sec.altermag}

While the Rashba effect is due to SOC, a non-relativistic spin splitting, termed altermagnetism~\cite{PhysRevX.12.040501}, is found in antiferromagnets of certain symmetry, where atomic sublattices carrying opposite spin are related by rotations. %
A higher order effect in which gradients of curvature (but not a constant curvature) induce a magnetization was recently predicted in this class of materials~\cite{PhysRevLett.134.116701}. %
Relatedly, an unconventional form of odd-parity or p-wave magnetism with non-relativistic spin-splitting has been predicted in noncollinear antiferromagnets with inversion symmetry breaking~\cite{hellenes2024pwavemagnets}. Because curvature breaks inversion symmetry, it offers a route to $p$-wave magnetism in otherwise centrosymmetric, collinear magnets, e.g. by rolling two-dimensional layers into nanotubes, making the spin texture non-collinear~\cite{jin2026rolling2d_pwave}.

\section{Outlook and Conclusions}\label{sec.concl}

The common thread of this chapter is that a strain gradient is a parity-odd field written into the lattice itself: sourced by geometry, growth or patterning rather than applied from outside, confined with a locality that stray fields cannot match, and, at the nanoscale and in two-dimensional crystals in particular, often unavoidable. What it couples to is dictated by symmetry alone (Table~\ref{table.sym}). As a time-even polar tensor it polarizes every insulator, whereas its coupling to magnetization requires magnetic order and is confined to the magnetic point groups that host the linear magnetoelectric effect, with a handful of additions. Universality is nevertheless recovered in the magnetic case through the inversion-symmetry breaking that curvature imposes on every magnet and the Dzyaloshinskii--Moriya interaction it necessarily brings with it.

For flexoelectricity the theory has reached maturity. The puzzles that long obscured the bulk nature of the effect, surface piezoelectricity and the static-versus-dynamic character of the coefficients, have been resolved, and the long-wavelength extension of density-functional perturbation theory~\cite{PhysRevX.9.021050,PhysRevB.105.064101} delivers the full flexoelectric tensor of an arbitrary crystal, within publicly available codes~\cite{abinit2020,abinit2025}. Two-dimensional crystals are the setting in which these coefficients are well-defined static materials properties, and where the low bending modulus makes curvature experimentally accessible~\cite{PhysRevLett.127.216801,PhysRevLett.131.236203}. The open questions are now largely experimental: a quantitative confrontation of first-principles coefficients with measurements on suspended monolayers, and the use of the anisotropic in-plane response to engineer polarization textures with non-trivial topology.

Flexomagnetism is at an earlier stage. Its symmetry properties are established, candidate materials are identified among the linear magnetoelectric antiferromagnets and their two-dimensional counterparts~\cite{Mangeri_2026,swjl-y8z1,PhysRevB.109.014410}, and the first experiments on strain-graded Cr$_2$O$_3$~\cite{Makushko2022} and La$_{0.67}$Sr$_{0.33}$MnO$_3$~\cite{PhysRevLett.127.117204} films provided evidence for the linear magnetization response and the curvature-induced DMI, respectively. What is missing is a predictive theory at the level flexoelectricity now enjoys. Secs.~\ref{sec.intrinsicDMI} and~\ref{sec.nonlin} show why this matters: the intrinsic curvature-induced corrections to the magnetic parameters are of the same order as the geometric terms of curvilinear magnetism~\cite{PhysRevLett.128.177202}, they introduce qualitatively new couplings such as a curvature-induced rotation of the easy axis, and some of their components cannot be extracted from any cylindrical geometry. Supercell calculations cannot close this gap systematically. Generalizing the long-wavelength perturbative framework to noncollinear spins, building on recent advances in the treatment of coupled spin--lattice response~\cite{Royo-26,Stengel-26}, or alternative schemes such as cyclic density-functional theory~\cite{BANERJEE2016605}, would allow flexomagnetic tensors to be computed on the primitive cell like their electric counterparts, and appears to us the most important methodological goal in the field.

The enabling experimental technologies are already in place. Freestanding
oxide membranes~\cite{Cai2022} and curved two-dimensional magnets~\cite{Caha2025} sustain
strain gradients of $10^{6}$--$10^{7}$~m$^{-1}$ or more, while three-dimensional
nanomagnetism~\cite{Pacheco2017,Donnelly2017,Donnelly2022,Gubbiotti_2025,Sabri2025}
and nano-sculpting of single-crystal magnets~\cite{Birch2026}
provide curvatures of similar magnitude in controlled geometries, in which
chiral~\cite{SanzHernandez2020,Farinha2025} and
topological~\cite{Volkov2024,Bezsmertna2024} spin textures are engineered by
shape rather than chemistry. 
Bent crystals, in which the lattice itself
carries the strain gradient, are the natural testing ground for the intrinsic
flexo-couplings discussed above, and in magnetoelectric materials the flexoelectric and flexomagnetic responses are not independent: the flexoelectric polarization is itself a source of magnetization, and the curvature-induced Dzyaloshinskii--Moriya interaction has an electric-field-induced twin. Flexomagnetoelectricity~\cite{Shen2018,PhysRevLett.133.156505}, the joint response of polarization and magnetization to strain gradients, has so far been addressed in only a few works and offers a rich playground for new    phenomena and functionality at the nanoscale.

\section*{Acknowledgements}
A.E. acknowledges financial support from the Swedish Research Council (VR 2022-04720), \AA Forsk (22-441), the G\"oran Gustafsson Foundation, the Swedish e-Science Research Center (SeRC), and The Wallenberg Initiative Materials Science for Sustainability (WISE) funded by the KAW Foundation. 
M.S. was supported by the Spanish MCIN/AEI/10.13039/501100011033 through grant PID2023-152710NB-I00 and through a Severo Ochoa Excellence award to ICMAB-CSIC, Grant CEX2023-001263-S.

\bibliographystyle{ieeetr}
\bibliography{references.bib}

\end{document}